\documentclass[prb,superscriptaddress,floatfix,twocolumn]{revtex4}
\usepackage{graphicx}

\def\lsco{La$_{2-x}$Sr$_x$CuO$_4$}
\def\lbco{La$_{2-x}$Ba$_x$CuO$_4$}
\def\lbcoate{La$_{1.875}$Ba$_{0.125}$CuO$_4$}
\def\lnsco{La$_{1.6-x}$Nd$_{0.4}$Sr$_x$CuO$_4$}
\def\lesco{La$_{1.8-x}$Eu$_{0.2}$Sr$_x$CuO$_4$}
\def\ybco{YBa$_2$Cu$_3$O$_{6+x}$}

\def\bscco{Bi$_2$Sr$_2$CaCu$_2$O$_{8+\delta}$}

\def\tco{$T_{\rm co}$}

\begin{document}

\title{Evidence for unusual superconducting correlations coexisting with stripe order in La$_{1.875}$Ba$_{0.125}$CuO$_4$}
\author{J. M. Tranquada}
\author{G. D. Gu}
\author{M. H\"ucker}
\author{Q. Jie}
\affiliation{Condensed Matter Physics \&\ Materials Science Department, Brookhaven National Laboratory, Upton, NY 11973-5000}
\author{H.-J. Kang}
\altaffiliation{New address: Dept.\ of Physics \&\ Astronomy, Clemson University, Clemson, SC 29634-0978}
\affiliation{NIST Center for Neutron Research, National Institute of Standards and Technology, Gaithersburg, Maryland 20899}
\author{R. Klingeler}
\affiliation{Leibniz-Institute for Solid State and Materials Research, IFW Dresden, 01171 Dresden, Germany}
\author{Q.~Li}
\affiliation{Condensed Matter Physics \&\ Materials Science Department, Brookhaven National Laboratory, Upton, NY 11973-5000}
\author{N. Tristan}
\affiliation{Leibniz-Institute for Solid State and Materials Research, IFW Dresden, 01171 Dresden, Germany}
\author{J. S. Wen}
\author{G. Y. Xu}
\author{Z. J. Xu}
\author{J. Zhou}
\affiliation{Condensed Matter Physics \&\ Materials Science Department, Brookhaven National Laboratory, Upton, NY 11973-5000}
\author{M. v. Zimmermann}
\affiliation{Hamburger Synchrotronstrahlungslabor (HASYLAB) at Deutsches Elektronensynchrotron (DESY), 22603 Hamburg, Germany} 
\date{\today}
\begin{abstract}
We present new x-ray and neutron scattering measurements of stripe order in La$_{1.875}$Ba$_{0.125}$CuO$_4$, along with low-field susceptibility, thermal conductivity, and specific heat data.  We compare these with previously reported results for resistivity and thermopower.  Temperature-dependent features indicating transitions (or crossovers) are correlated among the various experimental quantities.  Taking into account recent spectroscopic studies, we argue that the most likely interpretation of the complete collection of results is that an unusual form of two-dimensional superconducting correlations appears together with the onset of spin-stripe order.  Recent theoretical proposals for a sinusoidally-modulated superconducting state compatible with stripe order provide an intriguing explanation of our results and motivate further experimental tests.  We also discuss evidence for one-dimensional pairing correlations that appear together with the charge order.  With regard to the overall phenomenology, we consider the degree to which similar behavior may have been observed in other cuprates, and describe possible connections to various puzzling phenomena in cuprate superconductors.
\end{abstract}
\maketitle

\section{Introduction}

Many theorists have proposed that antiferromagnetic spin correlations should be relevant to the superconducting mechanism in layered cuprates.\cite{scal86,miya86,ande87,mont91,emer97,deml04,lee06,maie07,kanc08}  To the extent that they reflect the character of the Mott-insulating parent compounds, the coexistence of these antiferromagnetic correlations with metallic conductivity in underdoped cuprates requires concepts that go beyond conventional Fermi liquid theory.  One solution that has been proposed is the formation of hole-rich stripes separating narrow antiferromagnetic domains.\cite{zaan01,mach89,kive03,sach91,cast95,whit98a}  Static stripe order has been experimentally verified in a few special cuprate compounds, such as \lbco\ (LBCO),\cite{fuji04,abba05,kim08,duns08} \lnsco\ (Refs.~\onlinecite{tran95a,ichi00,chri07}), and \lesco\ (Refs.~\onlinecite{huck07,fink08}).   The stripe picture tends to draw comparisons with conventional charge- and spin-density-wave (CDW and SDW) systems, where the formation of density waves is associated with the gapping of electronic states near the Fermi level.  Such particle-hole gapping removes states that might otherwise contribute to the particle-particle gap that stabilizes the superconducting state.  Thus, to the extent that stripe order in cuprates is like conventional CDW/SDW order, one would expect it to compete with superconductivity.  Of course, the stripe order in cuprates does not develop as an instability of a Fermi liquid, but rather in response to doping a Mott insulator.  Nevertheless, the experimental observation that the bulk superconducting transition temperature, $T_c$, is a minimum when the stripe ordering temperature is a maximum\cite{tran97a,mood88} tends to reinforce objections to stripes being relevant to superconductivity.\cite{rice97,scal01,lee08}

One of the first indications that the story might be more complicated came with the observation that an unusual gap appears in the in-plane optical conductivity together with the onset of charge order.\cite{home06}  Next, a combination of photoemission and tunneling measurements provided evidence for a $d$-wave-like gap at low temperatures, within the stripe-ordered phase but above the bulk superconducting transition, $T_c$.\cite{vall06}  These signatures were quite suggestive of superconductivity, and they motivated a careful examination of transport and susceptibility measurements.\cite{li07}  The latter study provided evidence that two-dimensional (2D) superconducting correlations  coexist with stripe order in \lbcoate\ at temperatures as high as 40~K.  Thus, it appears that stripe order is quite compatible with pairing and superconductivity; however, something about the combined superconducting and stripe-ordered state frustrates the usual Josephson coupling between layers that one would expect to result in 3D superconductivity.  It has been proposed that a sinusoidally-modulated superconducting state, minimizing overlap with the spin order, in combination with the 90$^\circ$ rotation of the stripe orientation from one layer to the next,\cite{vonz98} can explain the frustrated Josephson coupling.\cite{hime02,berg07}  Independent analyses also indicate that the energy of  superconductivity coexisting with charge-stripe order is competitive with that of a uniform $d$-wave state.\cite{hime02,racz07,cape08,yang08,lee08b,whit08,berg08}  

While the theoretical work is encouraging, the claim of 2D superconductivity is still quite surprising.  If it is correct, then it has strong implications for the nature of superconductivity in the cuprates; thus, it deserves to be carefully examined.  In this direction, we present in this paper further characterizations of \lbcoate.

The rest of the paper is organized as follows.  The experimental methods are described in the next section, followed in Sec.~III by a presentation of results and comparison of transitions in various measured quantities.  An extended discussion is given in Sec.~IV.  There, we point out related observations in other cuprates, consider possible explanations in terms of spurious phases, review theoretical proposals for the antiphase superconducting state, provide further interpretations of the data, and discuss more general implications for understanding superconductivity in the cuprates.

\section{Experimental Details}

The crystals studied here were grown in an infrared image furnace by the floating-zone technique.  They are pieces from the same crystals used previously to characterize the optical conductivity,\cite{home06}
photoemission and STS,\cite{vall06} magnetization,\cite{huck05b} and magnetic excitations.\cite{tran04}  In particular, the charge-stripe order has been characterized previously by soft x-ray resonant diffraction.\cite{abba05} 

For the present study, the charge ordering and structural phase transitions have been measured using 100-keV x-rays on beam line BW5 at HASYLAB,\cite{vonz98,zimm08,bouc98} with the sample cooled by a displex closed-cycle refrigerator. Diffraction measurements were performed in transmission geometry with a beam spot size of $1\time1$~mm$^2$.  The experiment has been repeated several times with varying crystal orientation and a typical crystal thickness of 1~mm.

The spin-stripe ordering was determined by neutron diffraction measurements on the SPINS triple-axis spectrometer at the NIST Center for Neutron Research (NCNR).  The diffraction experiments were done with monochromator and analyzer crystals of pyrolytic graphite, using the PG(002) reflection to select 5-meV neutrons.  The effective horizontal collimations were $55'$--$80'$--$80'$--open.  For inelastic measurements, the final neutron energy was fixed at 5 meV, and a cooled Be filter was placed after the sample to minimize contamination from neutrons at harmonic wavelengths.  Again, the sample was cooled with a closed-cycle displex refrigerator.

The resistivity and thermopower results are taken from Ref.~\onlinecite{li07}, where the measurement techniques are described.  The magnetic susceptibility measurements were performed in a Quantum Design Magnetic Properties Measurement System (MPMS); compensation for remnant fields was used to enable accurate zero-field cooling.  The thermal conductivity was measured in a Quantum Design Physical Properties Measurement System (PPMS).  The specific heat measurements were performed in Dresden using another PPMS on a crystal with a mass of 35~mg.


\section{Results}

\subsection{Structural transitions}

The structural transition from the high-temperature tetragonal (HTT) phase (space group $I4/mmm$) to the low-temperature orthorhombic (LTO) phase ($Bmab$) occurs at $T_{d1}$.   The measurements of $S_{\rm ab}$ and $\kappa_{\rm ab}$ show jumps that indicate $T_{d1}=247$~K, while the x-ray study (on a different crystal) indicated $T_{d1}\approx235$~K, based on the point where the orthorhombic strain goes to zero.\cite{zimm08}  A laboratory x-ray study on a related piece of crystal reported $T_{d1}=232$~K.\cite{zhao07}  The transition temperature is certainly sensitive to composition, and we know that there is a slight variation in Ba concentration along the length of the floating-zone-grown crystals.  We have generally tried to study crystals taken from the late part of the growth, where the actual composition approaches the nominal one asymptotically, but we have not had perfect controls on this. Another factor could be sensitivity of $T_{d1}$ to crystal size and geometry, as strain effects lead to a twin-domain structure in the orthorhombic phase.  It might be of interest to systematically probe such behavior, but we have not yet done so.

\begin{figure}
\centerline{\includegraphics[width=3.2in]{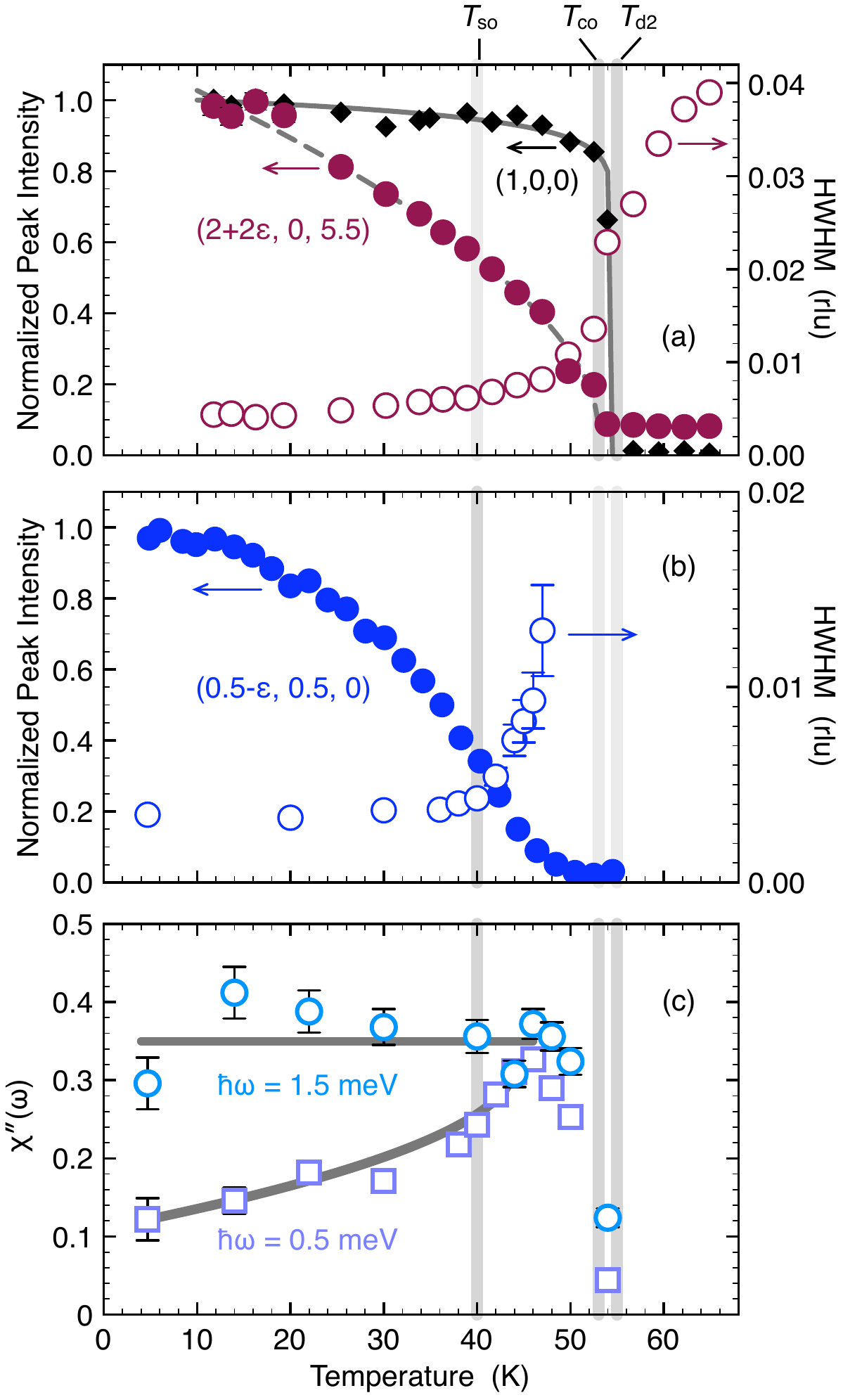}}
\caption{(color online) (a) Temperature dependence of normalized peak intensities of the charge-order peak $(2+2\epsilon,0,5.5)$ (filled circles) and the LTT superlattice peak (1,0,0) (diamonds) obtained by 100-keV x-ray diffraction; transverse width of the charge-order peak indicated by open circles.  Lines through the data points are guides to the eye.  (b) Normalized peak intensity (filled circles) and transverse width (open circles) of the magnetic peak $(0.5-\epsilon,0.5,0)$ measured by neutron diffraction with a neutron energy of 5 meV.  (c)  {\bf Q}-integrated $\chi''(\omega)$ for $\hbar\omega=0.5$~meV (squares) and 1.5 meV (circles).  Lines through the data are guides to the eye.  Vertical lines in all panels denote transition temperatures, as discussed in the text.}
\label{fg:diffract} 
\end{figure}

We are more interested in $T_{d2}$, the transition from LTO to the low-temperature tetragonal (LTT) phase ($P4_2/ncm$).  This can be detected by looking for the appearance of the (100) reflection, which is allowed in the LTT phase, but not in the LTO phase.  Figure~\ref{fg:diffract}(a) shows that $T_{d2}\approx 55$~K.  The transition has a finite width, as it is first-order at this composition.  The same transition temperature was observed in a synchrotron x-ray study by Kim {\it et al.},\cite{kim08} but Zhao {\it et al.}\cite{zhao07} found $T_{d2}=60$~K, similar to what Fujita {\it et al.}\cite{fuji04} originally found.  Like $T_{d1}$, $T_{d2}$ is sensitive to composition, shifting in the opposite direction to $T_{d1}$ with Ba concentration.  

For the discussion below, we will make use of coordinates corresponding to the HTT phase.  Thus, $a=3.78$~\AA, $c=13.2$~\AA, and wave vectors are expressed in units of $(a^*,a^*,c^*)$, with $a^*=2\pi/a=1.66$~rlu (rlu$=$ reciprocal lattice units).

\subsection{Charge order}

The onset of charge order was determined by measuring the $(2+2\epsilon,0,5.5)$ superlattice peak with x-rays.  The measured peak intensity and half-width-at-half-maximum (HWHM) obtained from scans along $(2+2\epsilon,k,5.5)$ are shown in Fig.~\ref{fg:diffract}(a).  (For examples of similar peak scans, see Ref.~\onlinecite{zimm08}.)   We identify the charge-ordering transition as $T_{\rm co}=53\pm1$ K, where the peak intensity stops decreasing and the peak width makes a sharp increase.  For these transverse scans, there is a weak and broad peak that remains for $T>T_{\rm co}$; however, this could be due to structural diffuse scattering.  Detailed scans along the modulation direction would be necessary to test whether the residual scattering might be related to quasi-static charge-stripe correlations.  Such scans at $T<T_{\rm co}$ show that $2\epsilon=0.24$.

$T_{\rm co}$ appears to correspond to the point at which the transition to the LTT phase is complete.  We note that the temperature dependence of the peak intensity is roughly consistent with that seen by Abbamonte {\it et al.}\cite{abba05} with soft-x-ray resonant diffraction; however, it is different from that observed by Kim {\it et al.}\cite{kim08} on a related crystal using x-rays of $\sim10$~keV. Kim and coworkers found the peak intensity to get quite weak by 40 K, and also found $2\epsilon=0.23$, slightly smaller than our result.   The cause of the discrepancy is unclear; however, one possibility involves differences in sensitivity to near-surface vs.\ bulk regions.  For example, the soft-x-ray resonant diffraction studies are extremely sensitive to surface preparation.  The positive detection of a charge-order peak was made on a cleaved crystal,\cite{abba05} whereas no peak was seen when a crystal with a polished surface was tried. The surface of the sample studied with $\sim10$-keV x-rays had been polished, and the penetration depth for such measurements is a few microns, in contrast to the 100-keV measurements where the diffraction is measured in transmission through a 1-mm thick crystal.  

Measurements of charge order peaks have been repeated several times on various crystals.  While the observed temperature dependence of the peak intensity has been consistent, there has been some variation of the peak width that seems to be beyond resolution effects.  Effective correlation lengths in the transverse direction (parallel to the stripes), obtained from the inverse of the HWHM, are in the range of 150--240~\AA\ at base temperature (12 K).

\subsection{Spin order and fluctuations}

The elastic neutron scattering results for magnetic peaks are similar to those obtained on \lnsco,\cite{tran99a} in that the width of the Lorentzian peak shape starts to grow before the peak intensity goes to zero [see Fig.~\ref{fg:diffract}(b)].  In the case of \lnsco, it was argued that the elastic signal detected at higher temperatures was due to integration over low-energy spin fluctuations.  The same argument applies here, as a spin-ordering temperature, $T_{\rm so}$, of 40~K has been identified by muon spin rotation ($\mu$SR) spectroscopy\cite{savi05} and by a single-crystal magnetization study\cite{huck05b}; this corresponds to the temperature at which the peak width starts to grow.   (An early $\mu$SR study on a polycrystalline sample found $T_{\rm so}=38$~K.\cite{luke91})  At low temperatures, the spin-spin correlation lengths are $\sim120$~\AA\ in the direction perpendicular to the stripes and $\sim600$~\AA\ parallel to the stripes.\cite{wen08b}

\begin{figure}
\centerline{\includegraphics[width=3.2in]{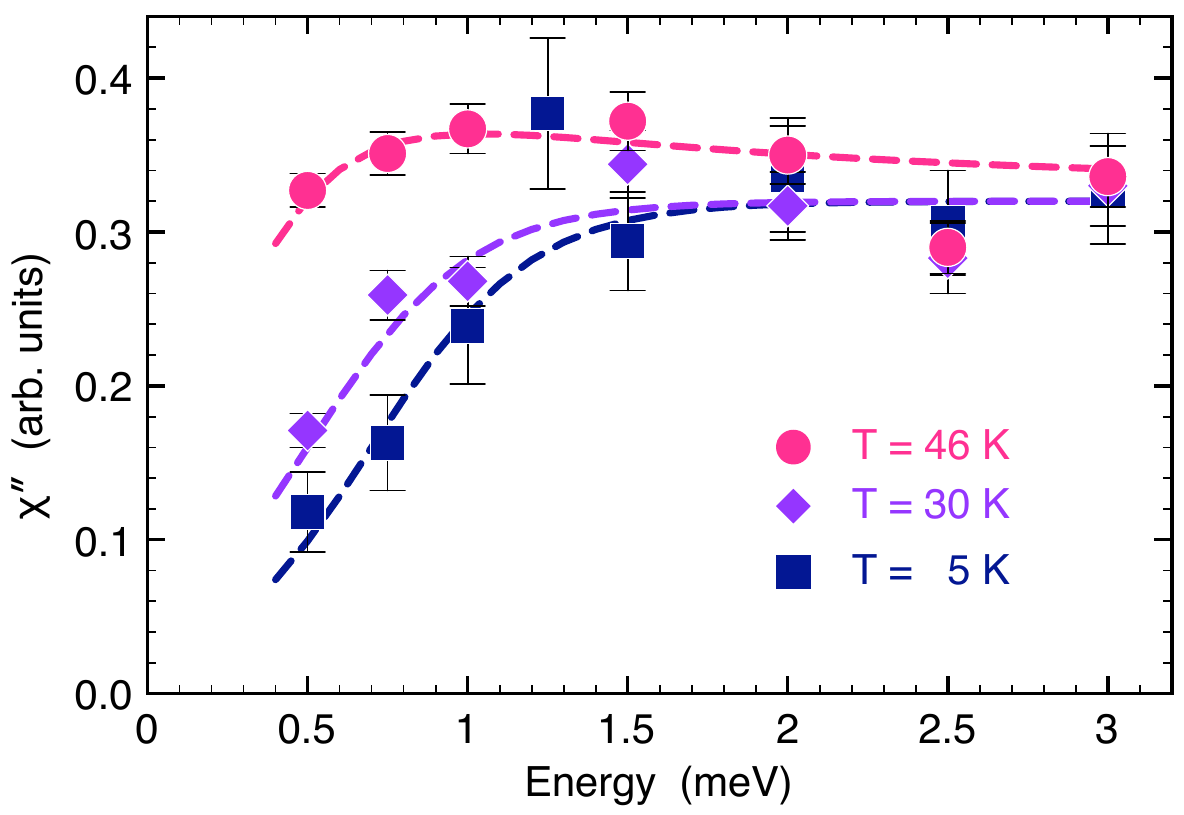}}
\caption{(color online) $Q$-integrated dynamic susceptibility vs.\ energy for incommensurate magnetic fluctuations.  Circles, diamonds, and squares represent results obtained at $T=46$~K, 30~K, and 5~K, respectively.  The dashed lines, which are effectively guides to the eye, are explained in the text.}
\label{fg:inel} 
\end{figure}

We have tested this scenario by directly measuring the low-energy fluctuations.  To do this, we made constant energy scans along ${\bf Q}=(\frac12+h,\frac12,0)$, through the peak at $h=0.12$, at various energies.    The peak width showed little variation with energy, so integration of the signal intensity over a one-dimesional scan should yield a result proportional to what one would obtain from a proper two-dimensional integration. We determined the integrated intensities and corrected for the Bose factor in order to obtain $\chi''(\omega)$.  The energy dependence of $\chi''(\omega)$ at several temperatures is shown in Fig.~\ref{fg:inel}.   At $T=46$~K, where the nominally elastic scattering is very weak, we find that $\chi''(\omega)$ is virtually independent of energy.  (A previous study has shown this to be the case in the 3--12 meV energy range.\cite{fuji04})  There appears to be a slight hump around 1.2 meV, and indications of a fall off below 0.5 meV.  For reference, the dashed line through the data corresponds to 
\begin{equation}
  \chi''(\omega) = 0.32\tanh(\hbar\omega/\Gamma) + 
     0.16{\hbar\omega\Gamma'\over (\hbar\omega)^2+\Gamma'^2},
\end{equation}
with $\Gamma=0.4$~meV and $\Gamma'=0.5$~meV; the energy resolution of the measurement should be $\sim0.4$~meV FWHM.
If we overlook the small variations with energy,  the general behavior is quite similar to what one would expect for spin waves in a 2D antiferromagnet, even though $\mu$SR and magnetization studies indicate the absence of order.  Thus, it appears that we have an effective spin-liquid state (spatially modulated by the charge-stripe order) at 46~K.  Note that our spin-liquid state is distinct from the much-discussed quantum spin liquid.\cite{lee08}

At lower temperatures, a gap develops.  The curves through the data in Fig.~\ref{fg:inel} correspond to
\begin{equation}
\chi''(\omega) = 0.16\left[ 1+\tanh\left({E-E_g\over\Gamma}\right)\right],
\end{equation}
with $E_g=0.5$ meV at 30~K and 0.7 meV at 5~K.   For more detailed characterization of the temperature dependence of the gap, we compare $\chi''$ at 0.5 meV and 1.5 meV vs.\ temperature in Fig.~\ref{fg:diffract}(c).   The values are comparable between $T_{\rm co}$ and $T_{\rm so}$, indicating a modulated spin-liquid state throughout this regime.  Near $T_{\rm so}$, $\chi''$(0.5~meV) starts to decrease, while $\chi''$(1.5~meV) stays roughly constant, indicating the opening of the gap.   
This gap is intriguing, since a spin gap frequently appears in the superconducting state, at least for optimal doping and above\cite{lake99,lee00}.  A spin gap associated with superconductivity (especially such a small one) should be reduced by an applied magnetic field.\cite{lake01,tran04b}  We tested this possibility in a separate experiment,\cite{wen08b} and found no significant change in $\chi''$(0.5~meV) due to application of a 7~T field at 30~K.  We conclude that the gap must be associated predominantly with spin anisotropy in the spin-ordered state.  This conclusion is consistent with the observation of a spin-flop transition at $H=6$~T.\cite{huck05b}  Furthermore, the observation of almost-gapless spin excitations for $T_{\rm so}<T<T_{\rm co}$ is consistent with appearance of anisotropy in the bulk susceptibility for $T<T_{\rm co}$.\cite{huck05b}

\subsection{Resistivity and magnetic susceptibility}

\begin{figure}
\centerline{\includegraphics[width=3.2in]{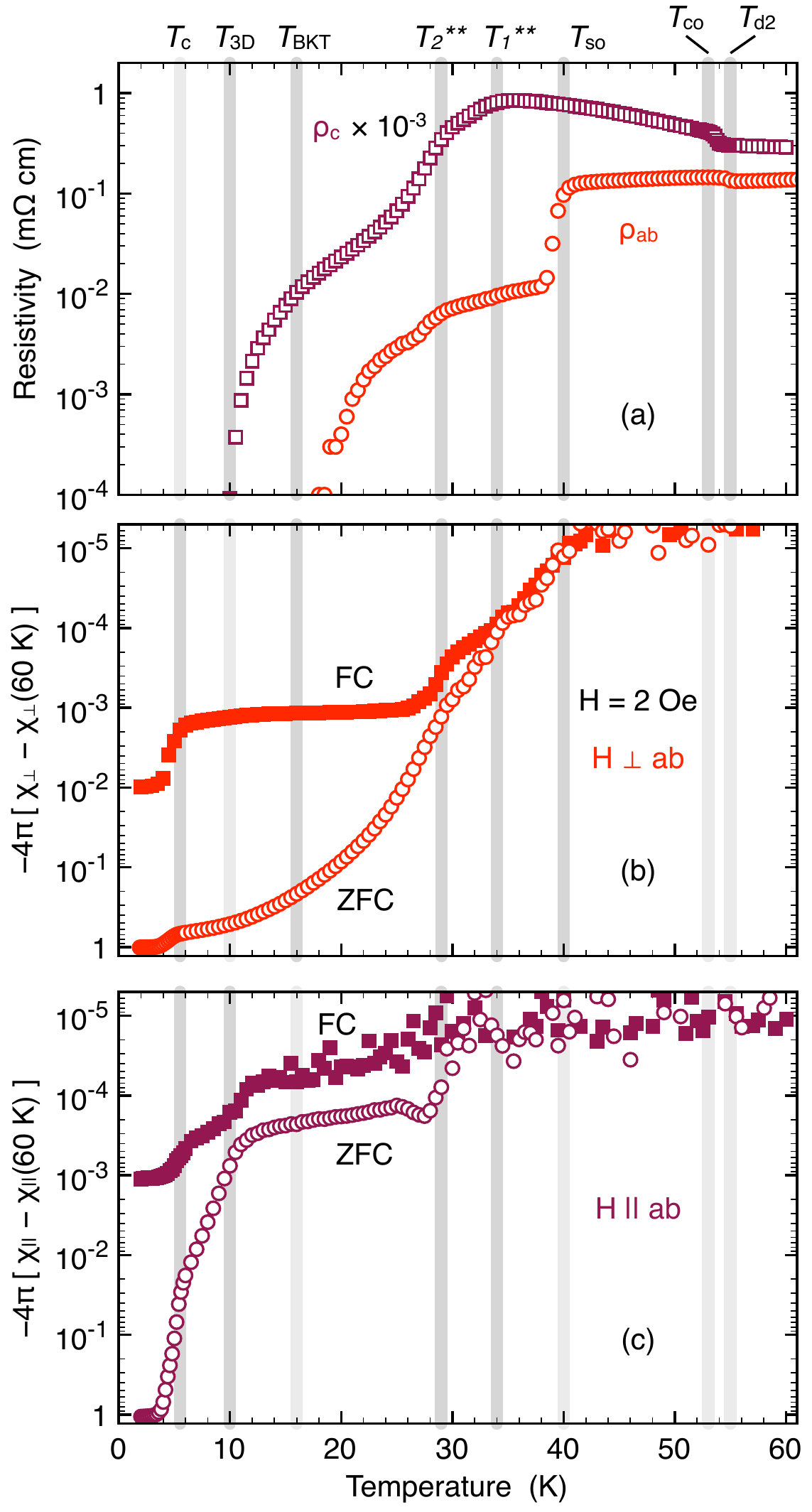}}
\caption{(color online) (a) In-plane resistivity, $\rho_{\rm ab}$ (circles), and resistivity perpendicular to the planes, $\rho_{\rm c}$, divided by $10^3$ (squares).  (b) Magnetic susceptibility measured with the field perpendicular to the planes, and (c) parallel to the planes, for an applied field of 2 Oe.  $\chi$ has been corrected for shape anisotropy, and the offsets (to allow plotting on a log scale) are: $4\pi\chi_\bot(60\ {\rm K})=4.2\times10^{-4}$ and $4\pi\chi_\|(60\ {\rm K})=1.6\times10^{-5}$.   Vertical gray lines denote relevant temperatures, with labels at the top.}
\label{fg:resist} 
\end{figure}

The resistivity data shown in Fig.~\ref{fg:resist}(a) are taken from Ref.~\onlinecite{li07}.  This particular plot emphasizes the fact that there is a sharp drop in $\rho_{\rm ab}$ at $T_{\rm so}$, but no corresponding change in $\rho_{\rm c}$.  New low-field susceptibility data, corrected for density and shape anisotropy, are shown in Fig.~\ref{fg:resist}(b,c); the normal-state susceptibility at 60 K has been subtracted to allow plotting on a logarithmic scale.  Results are shown both for field-cooling (FC) and zero-field-cooling (ZFC) measurements.  Panel (b) shows that a diamagnetic response appears at $T_{\rm so}$ when the magnetic field is applied perpendicular to the planes (inducing screening currents within the planes); there is no such diamagnetic onset when the field is parallel to the planes (requiring screening currents between planes).  These signatures are all consistent with the onset of 2D superconducting correlations at $T_{\rm so}$.

The c-axis resistivity just starts to turn down at $T_1^{\ast\ast}\approx34$~K, and drops more quickly below $T_2^{\ast\ast}\approx29$~K; there is a corresponding small step in $\rho_{\rm ab}$ at $T_2^{\ast\ast}$.  Comparing with the susceptibility, we see that $\chi_\bot$ becomes irreversible at $T_1^{\ast\ast}$ (FC and ZFC results diverge) and there is a small drop in the FC response at $T_2^{\ast\ast}$.  There is also an onset of a diamagnetic response in $\chi_\|$.  The combination of responses at $T_2^{\ast\ast}$ suggests that it represents the onset of superconductivity  in small 3D grains.  Clearly, these grains do not form a percolating path, as both $\rho_{\rm ab}$ and $\rho_{\rm c}$ remain finite after the grains become superconducting.  (Note that even at 10~K, the $\chi_\|$ ZFC response is still much less than 0.1\%\ of full shielding, consistent with a tiny 3D superconducting volume fraction.)  The FC $\chi_\bot$ response saturates below $T_2^{\ast\ast}$, suggesting that the 3D superconducting grains pin magnetic vortices.

At the apparent Berezinskii-Kosterlitz-Thouless transition, $T_{\rm BKT}\approx16$~K, discussed in Ref.~\onlinecite{li07}, $\rho_{\rm ab}$ goes to zero, while $\rho_{\rm c}$ remains finite.  (The identification of the transition as BKT-like is supported by nonlinear transport behavior reported in Ref.~\cite{li07}.)   The ZFC $\chi_\bot$ is already 20\%\ at this point, which is reasonable for the onset of 2D superconducting order throughout the sample.  The c-axis resistivity finally goes to zero at $T_{\rm 3D}\approx10$~K, where $\chi_\|$ starts to slowly decrease.   The final onset of bulk 3D superconductivity is $T_c\approx5.5$~K.  Note that the ZFC $\chi_\bot$ is already nearly 70\%\ of the full response at $T_c$.

\subsection{Thermoelectric power}

\begin{figure}
\centerline{\includegraphics[width=3.2in]{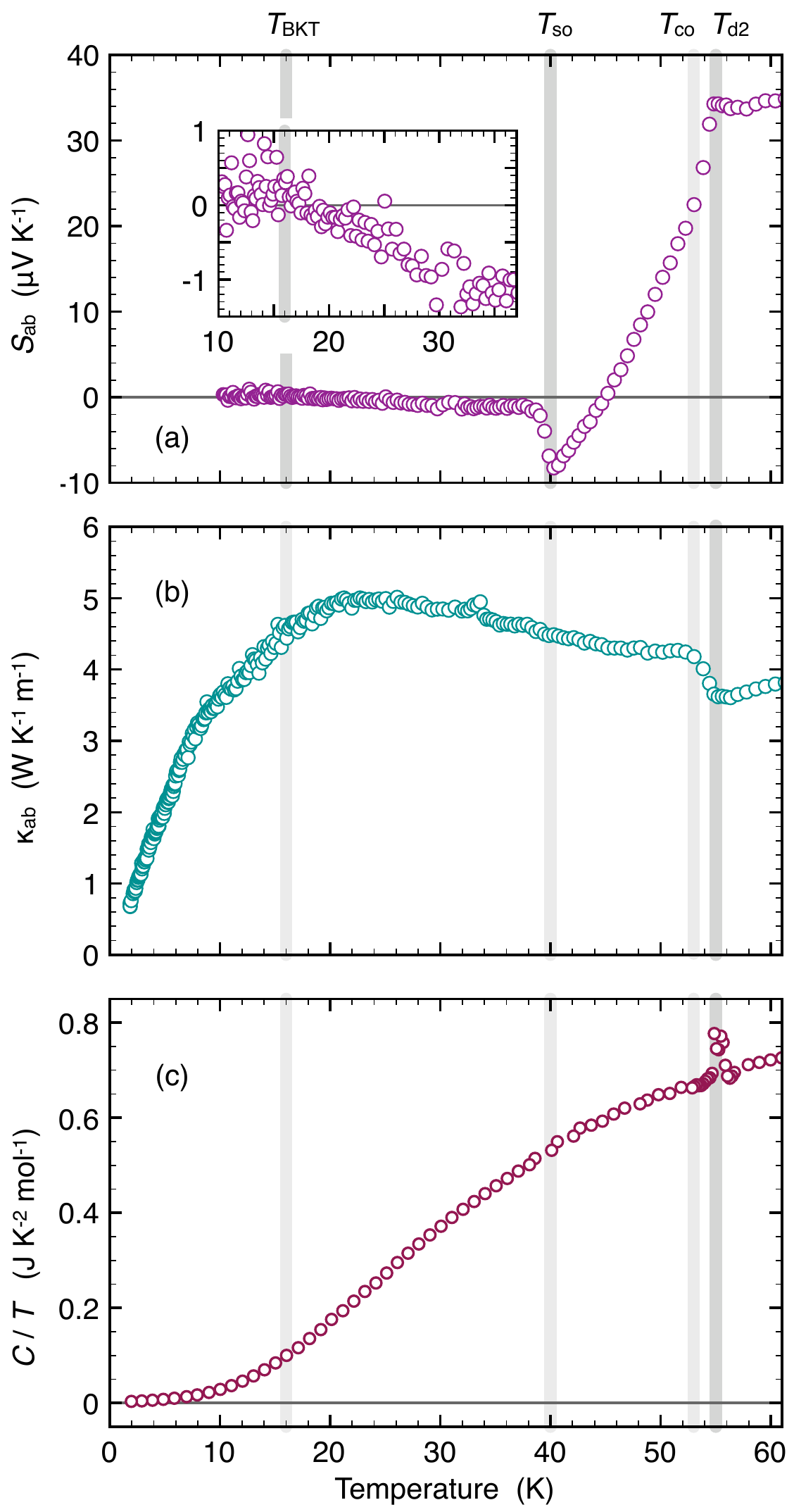}}
\caption{(color online) (a) Thermoelectric power measured parallel to the planes, $S_{\rm ab}$.  Inset shows portion of data on an expanded scale.  (b) Thermal conductivity, $\kappa_{\rm ab}$.  (c) Specific heat, $C$, divided by temperature.  Relevant temperatures are indicated by vertical gray lines.  }
\label{fg:sk} 
\end{figure}

The data for thermoelectric power measured parallel to the planes, $S_{\rm ab}$, shown in Fig.~\ref{fg:sk}(a) are taken from Ref.~\onlinecite{li07}.  Starting from the normal state, there is a sharp drop as soon as the structural transition at $T_{\rm d2}$ begins.  The thermopower continues to decrease with cooling, eventually going negative, before undergoing another drop in magnitude at $T_{\rm so}$.  The inset of Fig.~\ref{fg:sk}(a) gives an expanded version of $S_{\rm ab}$ below $T_{\rm so}$, showing that it remains finite down to $T_{\rm BKT}$; at the latter point, $S_{\rm ab}$ is essentially zero, consistent with superconducting order.

We have previously shown\cite{li07} that the jump in $S_{\rm ab}$ only matches up with $T_{\rm so}$ in zero magnetic field.  An applied field causes the jump to move to lower temperatures, following the jump in $\rho_{\rm ab}$.  In contrast, $T_{\rm so}$ increases slightly with field.

\subsection{Thermal conductivity}

The thermal conductivity measured parallel to the planes, $\kappa_{\rm ab}$, is shown in Fig.~\ref{fg:sk}(b).  It shows a sharp rise at $T_{\rm d2}$, and then reaches a maximum at $\sim21$~K, before decreasing at lower temperature.  Such behavior has been reported previously for La$_{1.88-y}R_y$Sr$_{0.12}$CuO$_4$ with $R=$ Nd and Eu\cite{babe98,sun03} and at the charge-stripe-ordering temperature in La$_{1.67}$Sr$_{0.33}$NiO$_4$.\cite{hess99}  As has been pointed out previously,\cite{babe98} this behavior is not seen in superconducting \lsco\ (LSCO), where uniform stripe order is not observed.  

On the other hand, the behavior of $\kappa_{\rm ab}$ also looks somewhat similar to what is observed in Bi$_2$Sr$_2$CaCu$_2$O$_8$ (Ref.~\onlinecite{kris97}) and \ybco\ (Ref.~\onlinecite{zein99}), where there is a rise below $T_c$.  In the latter cases, much of the enhancement in $\kappa_{\rm ab}$ below $T_c$ disappears when a strong magnetic field is applied,\cite{kris97,zein99} and it has been argued that this contribution is electronic.\cite{hirs96,zein99} To test for such a contribution in our sample, $\kappa_{\rm ab}$ has also been measured in a magnetic field of 9~T; we have not plotted that result because it looks virtually identical to the zero-field data.  Thus, the jump in $\kappa_{\rm ab}$ is distinct from that observed at $T_c$ in some cuprate superconductors; nevertheless, it could reflect a significant electronic effect, considering the substantial change in thermopower at the same temperature.    Contributions could also come from a change in acoustic-phonon lifetime due to stripe ordering,\cite{babe98}  or from the development of low-energy spin fluctuations.  Magnetic contributions to thermal conductivity can be substantial,\cite{hess01} and we have seen here [Fig.~\ref{fg:diffract}(c)] and previously\cite{fuji04} that there is a substantial jump in the weight and lifetime of low-energy spin fluctuations at $T_{\rm co}$.

\subsection{Specific heat}

Early heat capacity measurements on sintered samples revealed a cusp in $C/T$ vs.\ $T$ at the LTO-LTT transition temperature\cite{kuma91b}; a sharp peak was found at the corresponding transition in \lnsco.\cite{take01}  We clearly observe a significant peak at the first-order structural transition, as illustrated in Fig.~\ref{fg:sk}(c).

\begin{figure}[t]
\centerline{\includegraphics[width=3.2in]{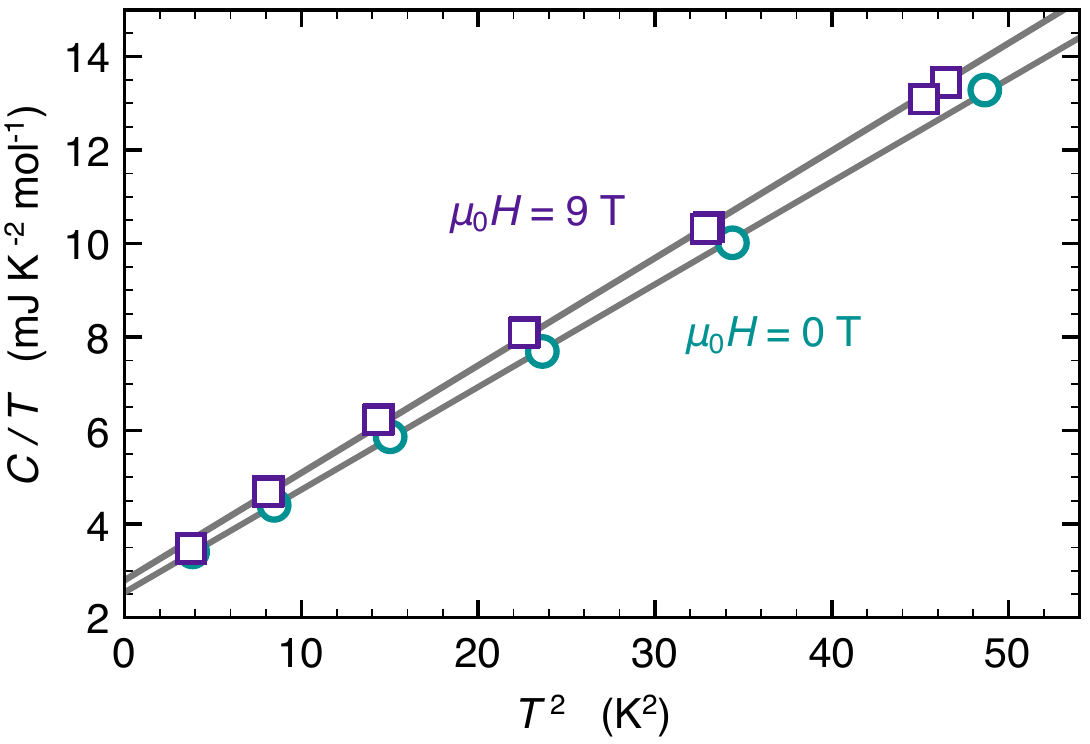}}
\caption{(color online) Specific heat divided by temperature,  plotted vs.\ $T^2$. Measurements in $c$-axis magnetic fields of 0 and 9~T represented by circles and squares, respectively.  Extrapolations of the fitted lines to $T=0$ provide estimates of $\gamma(T\rightarrow0)$ given in the text.}
\label{fg:spht} 
\end{figure}

Our results may not be of sufficient precision to identify changes in the electronic specific heat due to transitions below \tco; nevertheless, it is of interest to consider the electronic contribution in the low-temperature limit.  It has been observed previously\cite{okaj90} that the electronic specific heat coefficient $\gamma$ in LBCO is depressed for $x\sim\frac18$.  Partial substitution of Zn for Cu causes an increase in $\gamma$ that nearly saturates for 6\%\ Zn.\cite{aneg01}  Without Zn, $\gamma(T\rightarrow0)$ is measured to be 2-3 mJ K$^{-2}$ mol$^{-1}$, while an extrapolation of the results at high Zn-doping suggests that a ``normal'' state value would be $\sim11$ mJ K$^{-2}$ mol$^{-1}$.

The value of $\gamma(T\rightarrow0)$ can be obtained from a plot of $C/T$ vs.\ $T^2$, as shown in Fig.~\ref{fg:spht}.  In zero field, we find $\gamma(T\rightarrow0) = 2.5$ mJ K$^{-2}$ mol$^{-1}$.  Applying a $c$-axis magnetic field of 9~T increases $\gamma$ to 2.8 mJ K$^{-2}$ mol$^{-1}$.   Such a field-induced increase in $\gamma$ is commonly observed in $d$-wave superconductors such as LSCO.\cite{chen98,wang08}  The increase for our sample is about a third of that found in LSCO for the same field and doping level.

Note that the increase of $\gamma$ in a field is not consistent with an SDW interpretation of the depressed density of states.  We have observed that an applied magnetic field causes a slight increase in the spin stripe order parameter.  If the stripe order caused a particle-hole gapping of the Fermi surface, then the magnetic field should be increasing the gap and decreasing the electronic density of states (and $\gamma$), opposite to the experimental observations.

\section{Discussion}

\subsection{Is LBCO at $x=\frac18$ unique?}

There are several obvious questions  one can ask.  Given such striking and anomalous behavior, why have these features not been noticed previously?  Is it possible that LBCO at $x=\frac18$ is unique?  Is there any evidence for similar behaviour in related cuprates?

A two-step transition in the resistivity, starting at about 30~K, was observed in the initial studies of LBCO with $x=0.12$, using polycrystalline samples.\cite{mood88,sera89,maen91}  A weak diamagnetic response was also observed, starting near 30~K, that suggested only a few percent volume fraction of superconductivity.\cite{mood88}  These features were explained away as ``filamentary'' superconductivity, as there needed to be superconducting paths across the sample in order to explain the zero-resistivity state found at $\sim7$~K, where the diamagnetism was still quite small.  In fact, we initially adopted this interpretation when we observed the in-plane resistive drop at 40~K in a single crystal.\cite{home06}

Observations of the crucial anisotropy of the magnetic susceptibility and the resistivity had to await the growth of single crystals.  In the case of La$_{2-x-y}$Nd$_y$Sr$_x$CuO$_4$, where crystals first became available, the drop in resistivity for $x=0.12$ and $y\agt0.2$ occurs at a relatively low temperature and does not depend much on the direction in which the current flows; however, the resistive drop does tend to occur at a temperature significantly higher than the bulk onset of diamagnetism.\cite{naka92,saki99}  The diamagnetism can be a bit tricky to measure because of the large paramagnetic response of the Nd moments\cite{oste97}; nevertheless, Ding {\it et al.}\cite{ding08} have recently measured the anisotropy of the diamagnetic response in \lnsco\ with $x=0.10$, 0.15, and 0.18.  The results for $x=0.15$ look quite similar to those for LBCO, with a diamagnetic response in $\chi_\bot$, but not $\chi_\|$, starting at 21~K, followed by a 3D diamagnetic transition at $\sim13$~K, where $\rho_{\rm ab}$ effectively reaches zero.

A number of the anomalous in-plane properties that we report have been observed previously by others in related systems.  For example, the rapid drop in the thermoelectric power at $T\agt50$~K was found in polycrystalline LBCO for $0.10\alt x\alt0.13$ by Sera {\it et al.}\cite{sera89} and by Zhou and Goodenough\cite{zhou97}; Sera and coworkers also observed a related drop in the Hall coefficient, $R_{\rm H}$, for $x=0.10$ and 0.12.  The thermopower was found to go negative at lower temperatures for a narrow range of $x$ around $\frac18$.  In later work, it was shown that partial substitution of Nd for La in LBCO causes $T_{\rm d2}$ to increase but leaves the temperature dependence of the thermopower relatively unchanged.\cite{yama94b,bao97}  Related behavior was also reported for Nd- and Eu-doped LSCO.\cite{koik95,huck98}  The recent detection of charge-stripe order in La$_{1.8-x}$Eu$_{0.2}$Sr$_x$CuO$_4$ with $x=0.125$ and 0.15 ($T_{\rm co}\approx 80$~K and 70~K, respectively; $T_{\rm d2}=125$~K for both)\cite{fink08} confirms that the drop in $S_{\rm ab}$ is associated with the onset of charge order and is not intrinsic to the LTT phase.

After the discovery of stripe order in \lnsco, Noda {\it et al.}\cite{noda99} demonstrated that the previously observed drop in $R_{\rm H}$ also occurs in single crystals when measured with the current parallel to the planes and the magnetic field perpendicular.  The onset of the drop in $R_{\rm H}$ follows $T_{\rm co}$.\cite{ichi00}  Even more striking results were obtained for LBCO with $x=0.11$ by Adachi {\it et al.}\cite{adac01}   They showed that not only do $S_{\rm ab}$ and $R_{\rm H}$ both show abrupt drops at  $T_{\rm d2}$, but they both also go negative below 30~K.  Furthermore, when the temperature dependence of $R_{\rm H}$ is measured for different magnetic field strengths, $|R_{\rm H}|$ drops to zero at the temperature where $\rho_{\rm ab}$ reaches zero for the same field.\cite{adac01,adac05}  We assume that the same relationship between $R_{\rm H}$ and $S_{\rm ab}$ should hold at $x=\frac18$, though $R_{\rm H}$ has not yet been measured.  Similar field dependence of $R_{\rm H}$ on approaching $T_c$ has been observed\cite{suzu02} in La$_{1.84}$Y$_{0.04}$Sr$_{0.12}$CuO$_4$.  Without the yttrium in the crystal, $R_{\rm H}$ of La$_{1.88}$Sr$_{0.12}$CuO$_4$ does not go negative for magnetic fields of 6~T or less, but it does go negative at $T <18$~K for fields of 8~T and higher.\cite{suzu02}  (Note that a negative $R_{\rm H}$ was not observed in a study of LSCO films\cite{bala08}; however, this may be due to the sensitivity of stripe correlations to strain effects.)

The crossing of $R_{\rm H}$ from positive to negative on cooling, as seen in crystals with known stripe order, has also been observed in underdoped \ybco\ and YBa$_2$Cu$_4$O$_8$ samples that also exhibit quantum oscillations at sufficiently high magnetic fields.\cite{doir07,lebo07,bang08,jaud08,yell08}  These similarities are particularly clear if one compares Fig.~S1 in the supplementary material of Ref.~\onlinecite{lebo07} with results from Refs.~\onlinecite{adac01} and \onlinecite{suzu02}.   A couple of analyses of the quantum oscillations have already taken into account the possible relevance of stripe order.\cite{mill07,gran08}  We note that it may also be important to take into account the unusual correlations in the stripe-ordered phase that we will discuss further on.

Returning to single-layer cuprates, the suppression of interlayer superconducting phase coherence in the LTT structure was first demonstrated by Tajima {\it et al.}\cite{taji01}  They probed the Josephson coupling between superconducting cuprate layers by measuring infrared reflectivity with the polarization along the $c$ axis.  Studying La$_{1.85-y}$Nd$_y$Sr$_{0.15}$CuO$_4$, which changes its low-temperature structure from LTO to LTT as $y$ increases through $y_c=0.12$, they observed that the Josephson plasma edge shifted to lower energy and virtually disappeared as $y$ approached $y_c$ from below.  This rapid loss of interlayer coherence happens despite the fact that $T_c$, as determined from measurements of $\rho_c$, decreases rather gradually through $y_c$.

Using the same technique, Dordevic {\it et al.}\cite{dord03}  studied the Josephson plasma resonance (JPR) as a function of doping in LSCO.  They found that the resonance broadened for $x=\frac18$, indicating the presence of small domains with suppressed interlayer coherence.  Given the observation by Suzuki {\it et al.}\cite{suzu02} that $R_{\rm H}$ goes negative as a $c$-axis magnetic field exceeds 6~T, as mentioned above, one might expect to see an impact on the JPR.  Indeed, Schafgans {\it et al.}\cite{scha08} have now shown that the JPR is driven to zero at comparable fields and temperatures for the same \lsco\ compositions ($x=0.10$, 0.125) where field-induced magnetic order is found.\cite{kata00,lake02,chan08}

\subsection{What about spurious phases?}

\begin{figure}[t]
\centerline{\includegraphics[width=3.2in]{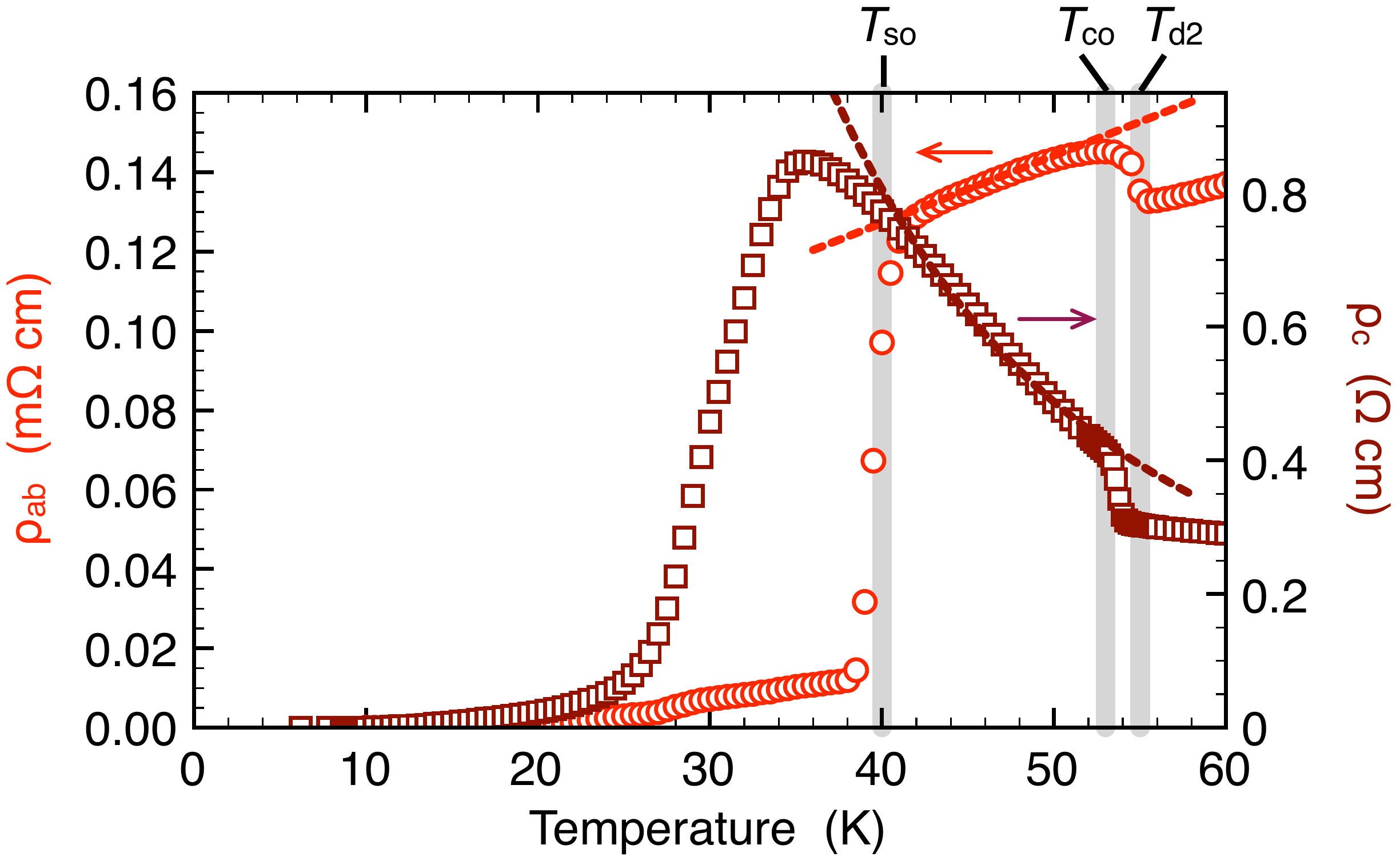}}
\caption{(color online) Temperature dependence of the planar and $c$-axis resistivity vs.\ temperature re-plotted on a linear scale.  The dashed lines correspond to $\rho_{\rm ab}=a_0+a_1T^{\alpha_{\rm ab}}$ with $\alpha_{\rm ab}=1$ and $a_0=0.059$ m$\Omega$ cm, and $\rho_{\rm c}=a_3T^{-\alpha_{\rm c}}$ with $\alpha_{\rm c}=2.25$. }
\label{fg:linear} 
\end{figure}

One of the arguments for the intrinsic nature of the superconducting correlations that onset at $T_{\rm so}$ is the observation that this temperature is well above the onset of bulk superconductivity\cite{mood88,sera89,abe99,adac05} for any composition in \lbco, where the highest transition temperature is $\sim32$~K (at $x=0.095$).  This certainly seems to rule out simple inhomogeneity in the local Ba concentration as an explanation.  Considering more exotic possibilities, it is known that compressive strain in thin films can raise the maximum $T_c$ of \lbco\ to 44~K (and adding excess oxygen to such films can raise it further to 49~K).\cite{sato00}  While the presence of such a strained phase as an impurity seems quite unlikely, given its apparent absence in crystals differing only slightly in Ba concentration,\cite{abe99,adac05} let us consider the possibility anyway.  To explain our observations, the impurity would have to be present in sufficient volume to give the large changes in $\rho_{\rm ab}$ [Fig.~\ref{fg:resist}(a)] and $S_{\rm ab}$ [Fig.~\ref{fg:sk}(a)].  At the same time, this impurity phase would have to have a geometry such that it would impact only the properties involving transport parallel to the CuO$_2$ planes of the dominant bulk material, with no hint of superconductivity for transport perpendicular to the planes.  To emphasize the strongly contrasting behavior between $\rho_{\rm ab}$ and $\rho_{\rm c}$ at $T_{\rm so}$, we plot the data on linear scales in Fig.~\ref{fg:linear}.  We cannot think of a credible way for such an extremely anisotropic response to be explained by an impurity phase.

For the sake of argument, let us consider the case of isolated, very thin layers extending across the sample in the directions perpendicular to the $c$ axis.  (For example, one might imagine some type of stacking fault in the LTT structure that allows stripes in neighboring planes to be parallel,\cite{kivepc} thus removing the frustration of the interlayer Josephson coupling.\cite{berg07})  The volume fraction of such layers would have to be quite small in order to be consistent with the observed resistivity anisotropy.  Suppose there were only one such layer, with a thickness of only a few unit cells (much less than the c-axis magnetic penetration depth).  This defect model could potentially explain the resistivity data; however, it would be inconsistent with the magnitude of $\chi_\bot$ measured by zero-field cooling, which is already quite substantial at $\sim20$~K.  A single superconducting layer would have a very weak effective diamagnetism because of a very large demagnetization factor.  On the other hand, if there were many very thin layers distributed along the $c$-axis direction, it would be more difficult to estimate the diamagnetic response.  

Of course, as already mentioned, we do have evidence for a very small volume fraction ($<0.1$\%) of a 3D superconducting phase that sets in at $T_2^{**}\approx29$~K, with possible onset behavior in $\rho_c$ at $T_1^{**}\approx34$~K.    While such an effective $T_c$ could potentially be explained by dopant inhomogeneity, it also corresponds to what one might expect to find if the structure were LTO, as in LSCO.  One possibility is that the finite correlation lengths of the spin and charge stripe orders limit the correlation length for superconducting order; such disorder might also relax the cancellation of Josephson couplings between neighboring layers. This could result in a 3D glass-like development of superconductivity.\cite{berg07}  One challenge for such a picture is the distinct evolution of $\chi_\|$ and $\chi_\bot$ at lower temperatures---one might expect to see more growth in the diamagnetism of $\chi_\|$ on further cooling, following $\chi_\bot$ in some fashion.   Another possibility would be the presence of a small volume fraction of LTO phase, as might be expected based on electron diffraction studies of LBCO with $x\sim0.125$ which suggest that the twin boundaries of the LTT phase have LTO-like character.\cite{zhu94,chen93b}  The twin boundaries typically extend as sheets perpendicular to the CuO$_2$ planes; however, the typical width within the planes might only be a few tens of {\AA}ngstroms.  Thus, there might be only select LTO domains that can support coherent superconductivity, which would be necessary in order to be consistent with the experimental fact that the 3D superconducting domains do not short out the resistivity.  In any case, the twin boundary spacing might provide an upper limit for the divergence of the phase coherence length of the 2D superconductivity.

Despite the weak 3D superconducting behavior, the growth of the diamagnetic shielding response in $\chi_\bot$ is remarkable.  The 2D diamagnetism is reversible between 40 K and $\sim30$~K, at which point the FC response seems to saturate, perhaps due to a flux-pinning effect associated with the 3D superconducting regions.  Nevertheless, the ZFC $\chi_\bot$ continues to grow, reaching 20\%\ of full shielding at the nominal $T_{\rm BKT}$.  This diamagnetic susceptibility is huge compared to the fluctuation diamagnetism typically seen in cuprates just above $T_c$,\cite{li94,li05} even for samples where an anomalous Nernst response is found to temperatures well above $T_c$.\cite{li07b,li07c} 

\subsection{Theoretical proposals}

All of the cuprate superconductors are dominated by 2D superconductivity; however, a finite Josephson coupling between the planes always leads to a transition to 3D superconductivity at a temperature higher than the nominal Berezinskii-Kosterlitz-Thouless temperature.  Thus, our observation of a BKT transition (inherently 2D behavior) in a 3D crystal is extremely unusual.  (It may not be unique, as nonlinear transport behavior consistent with a BKT transition at 85~K has been reported\cite{wan96} in slightly underdoped Bi$_2$Sr$_2$CaCu$_2$O$_{8+\delta}$.)  While this is an experimental surprise, it appears to be an example of a theoretically-predicted ``floating'' (or ``sliding'') phase.  O'Hern, Lubensky, and Toner\cite{oher99} first showed that, for a 3D stack of 2D $XY$ models coupled by current-current interactions, weak Josephson couplings are irrelevant (in the renormalization group sense) and the 2D layers behave as if they are decoupled.  That analysis applies directly to the case of a stack of 2D superconducting layers.  

The stripe problem is another one in which the relative coupling or decoupling of phases in parallel systems is of interest.  It was first proposed over a decade ago that metallic charge stripes in an antiferromagnetic background could develop pairing correlations.\cite{emer97,scal01}  It was assumed there,\cite{emer97} and in later analyses,\cite{kive98,vojt99,carl00} that the Josephson coupling between stripes would be ``in-phase'', leading, effectively, to uniform $d$-wave superconductivity.  It was also recognized that charge-density-wave (CDW) correlations within the charge stripes would compete with the superconducting instability.  Thus, it was anticipated that dynamic fluctuations or disorder of the stripes might be essential for achieving 2D superconductivity.\cite{kive98,vojt99}

Our experiments demonstrate that the static stripes in LBCO at $x=\frac18$ do not appear to have succumbed to CDW ordering.  There is, of course, an anomaly in the longitudinal bond-stretching phonon mode consistent with a strong electron-phonon coupling.\cite{rezn06}  This could well correspond to a coupling of bond-stretching vibrations to low-energy electronic excitations along a stripe; however, the wave vector of the anomaly corresponds to a nesting of the antinodal states at $4k_{\rm F}$, where $k_{\rm F}$ is a Fermi wave vector obtained from the photoemission results.\cite{vall06}  The softened phonon remains at a relatively high energy ($\sim60$~meV), so the coupling is purely dynamic.  Furthermore, the fact that the coupling is at $4k_{\rm F}$, rather than $2k_{\rm F}$, suggests that pairing correlations could be involved.

It turns out that this metallic stripe state does not violate theory.  It has been shown by Emery {\it et al.}\cite{emer00} that there is a range of parameter values describing the coupling of charge correlations in neighboring stripes for which it is possible to obtain a floating phase---in this case termed a ``smectic metal''---in addition to a smectic superconductor.  This result was developed further by Mukhopadhyay, Kane, and Lubensky,\cite{mukh01} who considered the case of a stack of stripe layers with the stripe direction rotating $90^\circ$ from one layer to the next, as occurs in our case.  They showed that it is possible to have a floating phase for which, in the limit $T\rightarrow 0$, one has $\rho_{\rm ab}\sim T^{\alpha_{\rm ab}}$ and $\rho_{c}\sim T^{-\alpha_{\rm c}}$, with both of the exponents, $\alpha_{\rm ab}$ and $\alpha_{\rm c}$, positive numbers.  We have a limited temperature range, 
$T_{\rm so} < T < T_{\rm co}$ (not all that close to the $T\rightarrow 0$ limit), in which we can compare with the prediction.  The dashed lines in Fig.~\ref{fg:linear} correspond to $\alpha_{\rm ab}=1$ and $\alpha_{\rm c}=2.25$.  We should note that one can extend the analysis to $T<T_{\rm so}$ by applying a $c$-axis magnetic field to suppress the superconductivity.  In this case, extrapolations of the power-law curves deviate quite a bit from the low-temperature resistivities.  The experimental $\rho_{\rm ab}$ develops upward curvature,\cite{li07} similar to the $\ln(1/T)$ behavior seen in underdoped LSCO and \lnsco\ at high fields and low temperatures.\cite{ando95,tran96b}  We also find that, while $\rho_{\rm c}$ continues to increase on cooling, the effective exponent $\alpha_{\rm c}$ decreases.

To understand our superconducting floating phase, we need a mechanism that would frustrate the Josephson coupling between neighboring layers.   A possible solution was first proposed by Himeda, Kato, and Ogata\cite{hime02} and then independently rediscovered by Berg {\it et al.}\cite{berg07}  The key idea is that the superconducting wave function oscillates sinusoidally, having a large magnitude on the charge stripes but changing sign from one stripe to the next.  Such an antiphase superconducting wave function goes to zero where the magnetic order is largest.  (In this sense, it is similar to the modulated superconducting state proposed by Fulde and Farrell\cite{fuld64} and by Larkin and Ovchinnikov\cite{lark64} (FFLO) in the case of coexisting superconductivity and ferromagnetism.)   When the antiphase superconducting state is combined with the orthogonal orientation of stripes in neighboring planes, it is not hard to see that the Josephson coupling between nearest-neighbor planes is frustrated.  (A discussion of possible longer-range Josephson couplings is given by Berg {\it et al.}\cite{berg07})

Several recent calculations have found that the antiphase superconductor, in combination with stripe order, is quite close in energy to the uniform $d$-wave state.\cite{hime02,racz07,cape08,yang08,lee08b,whit08,berg08}  Also, it has been noted that these distinct superconducting states will tend to compete with each other.\cite{agte08}  This has interesting implications that will be discussed below.

\subsection{Interpreting the data}

The temperature dependence of $\rho_{\rm ab}$ looks qualitatively similar to the results of Hebard and Vandenberg\cite{heba80} for a granular lead film.  In the latter case, the initial drop in resistivity was due to the onset of superconductivity within the lead grains, with the second transition corresponding to 2D superconductivity due to Josephson coupling between the grains.  Our situation has similarities and differences.  The drop in $\rho_{ab}$ at 40~K indicates the onset of 2D superconducting correlations.  To the extent that the phase-coherence length is finite, one might consider a region within a plane with a radius comparable to the phase-coherence length to be like a grain.  The second transition in each case involves achieving phase coherence throughout a plane/film.

In our case, the superconducting correlation length, $\xi_{\rm sc}(T)$, will be limited by topological defects, such as dislocations, in the stripe order.  If the correlation length for stripe order is determined by topological defects, then the stripe correlation length should provide an upper limit to $\xi_{\rm sc}$.\cite{berg07}  Of course, stripe correlations are also impacted by non-topological effects such as variations in stripe spacing,\cite{zach00,tran99a} leading to $\xi_{\rm so}>\xi_{\rm co}$ (as observed), so the stripe correlation length measured by scattering techniques might underestimate the limit on $\xi_{\rm sc}$.  Structural twin boundaries of the LTT phase are also likely to act as topological defects, so the width of a twin domain likely represents an ultimate limit to $\xi_{\rm sc}$.

While the change in $\rho_{\rm ab}$ at $T_{\rm so}$ is dramatic, there is an equally dramatic change in $S_{\rm ab}$ at the higher-temperature charge-ordering transition.  The drop in the in-plane thermopower signals a major change in the electronic density of states near the Fermi level.  The states that make the biggest contribution to the density of states are in the antinodal region of the Fermi surface.  Furthermore, a gap-like feature in the in-plane optical conductivity also appears at $T_{\rm co}$.\cite{home06}  Thus, it seems likely that the antinodal states become gapped with the onset of charge order.  

The increased magnitude of $d\rho_{\rm c}/dT$ for $T\alt T_{\rm co}$ is consistent with such a picture.   An analysis of electronic structure in the cuprates\cite{chak93,ande95} indicates that conduction along the $c$ axis should involve Cu $4s$ states, which couple to the planar conduction band only in the antinodal region.  It follows that gapping the antinodal states should cause a substantial increase in $\rho_c$.  

In contrast, $\rho_{\rm ab}$, which is determined largely by near-nodal states,\cite{lee05} changes very little at $T_{\rm co}$.  Thus, it appears that the nodal states are not greatly affected by the charge order, nor is there much impact from the slowing of the spin fluctuations to a virtually gapless state.   Note that the nodal wave vector is at 45$^\circ$ to the stripe direction, so that these states involve a 2D dispersion, whereas the antinodal states can be associated with states that disperse along the charge stripes.\cite{salk96,gran02,gran06,woll08}
 
To summarize, it appears that states associated with the charge stripes are gapped, but states that propagate at a significant angle to the stripes are relatively unaffected.  The nature of the gap in the optical conductivity does not seem to change on cooling through $T_{\rm so}$, which suggests that the gap that sets in at $T_{\rm co}$ is associated with pairing.  Thus, the data appear to be consistent with strong pairing correlations in the ordered charge stripes, but no coherence between neighboring stripes.  Such a  state appears similar to the smectic metal\cite{emer00} discussed above.
 
Previous studies of the temperature dependence of the spin fluctuations on warming through $T_{\rm co}$ have provided evidence for the presence of dynamic stripes in the LTO phase.\cite{fuji04,xu07}  Based on such evidence, there seems to be an electronic nematic phase\cite{kive98} present at $T>T_{\rm co}$.  An estimate of the critical temperature for the nematic phase is $\sim200$~K, the point at which the incommensurability of spin fluctuations can no longer be resolved for low excitation energies ($\sim3$~meV).\cite{fuji04}  If we intepret $T_{\rm co}$ as a transition from the nematic to the smectic phase (induced by electron-lattice coupling), then these results also have important implications for the nature of the nematic phase (and its differences from the smectic).  To the extent that the incommensurability of the low-energy spin fluctuations is driven by a spin gap on the hole-rich stripes, the spin gap is present in both phases; however, gapping of the charge properties only shows up in the nominal smectic phase.

On cooling through $T_{\rm so}$ (in zero field), we have a transition (or crossover) from the 1D pairing correlations of the smectic metal phase to the 2D superconducting correlations of a superconducting smectic phase.  The onset of local superconducting coherence, $T_c^{\rm 2D}$, appears to be limited by the development of spin-stripe order; as we have shown previously, $T_c^{\rm 2D}$ decreases rapidly as a $c$-axis magnetic field is applied,\cite{li07} whereas $T_{\rm so}$ is slightly enhanced by the applied field.\cite{savi05}  (Analysis of magnetic susceptibility indicates that $T_{\rm so}$ increases slightly with magnetic field oriented perpendicular to $c$.\cite{huck05b}) The neutron diffraction measurements alone do not prove that the spin-stripe order occurs uniformly within the planes; however, confirmation of uniform stripe ordering is provided by $\mu$SR\cite{nach98,savi02} and nuclear-quadrupole-resonance\cite{hunt01} studies.

We have inferred two steps in the development of the 2D superconducting correlations: first, pairing correlations develop within charge stripes, involving antinodal states, and then local 2D phase coherence develops.  (Related ideas have been presented previously by Perali {\it et al.}\cite{pera00} and by Tsvelik and Chubukov.\cite{tsve07})  The changes observed in $\rho_{\rm ab}$ are consistent with the idea that the near nodal states are essential for establishing the phase coherence.\cite{ido99,vall06b}  This behavior also seems consistent with the two-gap behavior that has been identified in underdoped cuprates,\cite{leta06,lee07} in which the $d$-wave-like gap in the near-nodal region can extrapolate to a smaller energy than the gap observed for the antinodal states.  It would be quite interesting to see with photoemission whether a gapless nodal arc\cite{norm98,kani06} of electronic states is present at the Fermi level for $T>T_{\rm so}$.

\subsection{Implications}

It is interesting to consider an older calculation by Monthoux and Scalapino.\cite{mont94}  With the assumption that pairing is due to the exchange of spin fluctuations, they considered superconductivity in a 2D Hubbard model and how the distribution of weight in the dynamic spin susceptibility should be adjusted to optimize $T_c$.   They found that antiferromagnetic spin fluctuations at energies of $\sim10kT_c$ are the best, but that low-energy spin fluctuations are bad for superconductivity.  The latter result seems qualitatively consistent with the idea that static spin order may be compatible with a modulated superconducting state---the superconducting wave function attempts to minimize its overlap with the troublesome spin correlations.  

This also has implications for superconducting LSCO, where uniform $d$-wave superconductivity has been shown to dominate.\cite{tsue04}   As one cools through $T_c$, it may be favorable to gap the low-energy spin fluctuations.  This is what appears to happen in LSCO from optimal to overdoping, with the weight from below the spin gap being pushed up to energies just above the gap.\cite{lake99,lee00,chri04,tran04b}  It is interesting that the size of the spin gap is $\sim2kT_c$ at optimal doping.  

One possibility is that the amount of low-energy magnetic spectral weight determines which type of superconductivity develops.  If the low-energy spin fluctuations are not too strong, then it may pay to spend some energy to gap these so that a uniform $d$-wave state can develop.  On the other hand, if there is a lot of low-energy, spatially-modulated spectral weight, with the extreme case being the presence of spin stripe order, then the antiphase superconducting state may win out.  These two different superconducting orders will compete with each other, but may also coexist.\cite{agte08}  With this possibility in mind, we consider the case of underdoped LSCO.  Some degree of static magnetic order has been observed by $\mu$SR\cite{nied98} out to a hole concentration of at least 10\%, and neutron scattering studies\cite{lee00,chan08,kofu08} show that there is no true spin gap in the superconducting state for $x\alt0.13$.  Furthermore, the ratio of the superfluid density to the normal-state carrier density is observed\cite{uemu89,taji05} to be low in LSCO compared to other cuprates such as \ybco, and a study\cite{omor07} of the magnetic susceptibility of single crystals indicates that the Meissner fraction decreases with $x$ for $x\alt0.09$.   Could these phenomena be associated with coexisting superconducting states?

Another relevant system is La$_2$CuO$_{4.11}$, where the interstitial oxygens are positioned in every fourth La$_2$O$_2$ layer.\cite{lee99}  Here spin-stripe order and bulk superconductivity appear simultaneously at $T_c=42$~K.  Applying a $c$-axis magnetic field of 7.5~T causes a slight increase of the magnetic ordering temperature but a significant decrease in $T_c$.\cite{khay02}  These behaviors are very similar to what we find in \lbco\ with $x=\frac18$, except for the fact that it is bulk superconductivity that appears at 42~K rather than 2D correlations.  The difference in phase coherence could be due to the absence of a crystal potential that could pin stripes in orthogonal directions in neighboring layers.  La$_2$CuO$_{4.11}$ has a different idiosyncrasy: it has one pair of CuO$_2$ layers that surrounds the interstitial oxygen layer and another pair that is further removed.\cite{note1}\nocite{well97,xion96}  The variation of the magnetic scattering with momentum perpendicular to the planes\cite{lee99} is consistent with correlated spin order in pairs of planes\cite{note2}; which of the inequivalent pairs might have the dominant spin order is not determined by experiment.   The fraction of Cu sites participating in magnetic order is estimated to be $\sim40$\%\ from $\mu$SR meausurements,\cite{savi02} while application of a $c$-axis magnetic field of 8~T increases the intensity of the magnetic superlattice peaks by $\sim50$\%.\cite{khay02}   Presumably there is some superfluid density in all layers.\cite{savi02}

In a related direction, Yuli {\it et al.}\cite{yuli08} have studied the enhancement of $T_c$ for thin films of LSCO by adding a highly-overdoped layer of La$_{1.65}$Sr$_{0.35}$CuO$_4$.   Interestingly, they find that the bilayer $T_c$ is highest for a base layer with $x=0.12$, even though there is a local minimum at that composition for the bare layers.  Their results are consistent with the idea that pairing is strong, and perhaps maximized, at $x\approx\frac18$ (as suggested by photoemission measurements\cite{vall06}), and that $T_c$ of the bare LSCO films are limited by phase fluctuations.  

More generally for underdoped cuprates, many groups have considered the idea that the ``normal'' state within a magnetic vortex core might actually involve an ordered state that competes with uniform $d$-wave superconductivity.\cite{arov97,deml01,wang01a,zhu02a,fran02,chen02,kive02,lee04}  The antiphase superconducting state is certainly a good candidate for this,\cite{agte08} and would provide a natural interpretation for the density-of-states modulations that have been detected within vortex cores of \bscco.\cite{hoff02,levy05}  It also makes sense for the field-induced incommensurate magnetism in LSCO.\cite{lake02,khay05,chan08}

Returning to the idea that low-energy spin fluctuations must be gapped in order to establish a uniform $d$-wave state, it could be that the spectral weight present at $T\agt T_c$ limits $T_c$.  The upper limit on pairing is generally considered to be the antinodal gap measured by angle-resolved photoemission spectroscopy (ARPES).   The antinodal gap is as large as 20 meV in LBCO and LSCO (Ref.~\onlinecite{vall06}), which is about 2/3 of the gap in \ybco\ and \bscco; thus, the antinodal gap does not seem to be limiting $T_c$ in LSCO.  Rather, there is more low-energy magnetic spectral weight in LSCO than in \ybco, and the spin gap that develops in the superconducting state for La$_{1.84}$Sr$_{0.16}$CuO$_4$ (Ref.~\onlinecite{lake99}) is $\sim1/4$ that in YBa$_2$Cu$_3$O$_{6.85}$ (Ref.~\onlinecite{bour00}). 

\acknowledgments

We are grateful to S. A. Kivelson, E. Fradkin, V. Oganesyan, T. M. Rice, M. Strongin, and A. Tsvelik for valuable discussions.  We acknowledge the support of the National Institute of Standards and Technology, U.S. Department of Commerce, in providing the neutron research facilities used in this work.  This work was supported by the Office of Science, U.S. Department of Energy under
Contract No.\ DE-AC02-98CH10886.


\end{document}